\documentstyle[12pt,aps]{revtex}
\begin{document}
\baselineskip 24pt plus2pt
\title {Transmission and reflection coefficients for a 
scalar field inside a charged black hole}
\author{Amos Ori\\
Department of Physics, Technion---Israel Institute of 
Technology\\
32000 Haifa, Israel} 
\date{\today}
\maketitle

\begin{abstract}

Using the late-time expansion we calculate the leading-order 
coefficients describing the evolution of a massless scalar 
field inside a Reissner-Nordstrom black hole. These 
coefficients may be interpreted as the reflection and 
transmission coefficients for scalar-field modes propagating 
from the event horizon to the Cauchy horizon. Our results agree 
with those obtained previously by Gursel et al by a different 
method.
\end{abstract}




\newpage
In a recent paper \cite{1}, we analysed the evolution of a massless 
scalar field inside a Reissner-Nordstrom black hole. The analysis 
was based on a new method, the {\it late-time expansion}, which is 
essentially an expansion in the small parameter $1/t$. This 
method takes advantage of the relatively simple form of the 
ingoing radiation tails, which are known to obey an inverse-power 
law. \cite{2} Assuming initial data $\psi \cong v^{-n}$ 
at the event horizon (EH), the asymptotic behavior near the 
Cauchy horizon (CH) was found in Ref. \cite{1} to be
\begin{eqnarray}
\psi =\sum\limits_{j=0}^\infty  {\,\left[ {a_j\,u^{-n-
j}\;\,+\,b_j\,v^{-n-j}} \right]+O\,(\delta r)}\;,
\label{1}
\end{eqnarray}
where $\delta r\equiv r-r_-$, $u$ and $v$ are the double-null 
Eddington-Finkelstein-like coordinates, and $a_j$ and $b_j$
 are constants (the notation here is the same as that of Ref. 
\cite{1}, except for a few changes specified below). Of special 
importance are the coefficients $a_0$
 and $b_0$, which dominate the evolution near the CH at late time (i.e. 
$v\to \infty \,\kern 1pt ,\,\kern 1pt u\to -\infty $).
These parameters may be interpreted (in a somewhat vague 
sense) as the reflection and transmission coefficients, 
respectively, for low-frequency scalar-field modes propagating 
from the EH to the CH. The explicit value of these coefficients 
(like all other coefficients $a_j$
 and $b_j$)
 was not calculated in Ref. \cite{1}. In this paper we shall use 
the late-time expansion to calculate $a_0$
 and $b_0$.
 The problem of a massless scalar field inside a charged black 
hole was previously analysed \cite{3,4,5} by a different method 
(Fourier integration in the complex plain), and the 
transmission and reflection coefficients were calculated in 
Ref. \cite{4}. Our goal here is to compare the results of the late-
time expansion to those obtained in Ref. \cite{4} by the other 
method.
	
We shall use here the same notation as in Ref. \cite{1}, except 
for the following changes:
\newline
(i) We denote here the "tortoise" coordinate by $r^*$ (not $x$);
\newline
(ii) $\kappa_-$, the inner horizon's surface gravity, 
is defined here in the standard way, 
$\kappa _- =(r_+-r_-)/(2r_-^2)$;
\newline
(iii) We denote the total scalar field by $\Psi $
(we reserve the symbol $\psi $
 for the spherical-harmonics modes of the field). $\Psi $
 satisfies the Klein-Gordon (KG) equation, $\Psi _{\;;\alpha 
}^{\kern 1pt ;\alpha }=0$.
	
We first decompose $\Psi $
 into spherical harmonics: 
$$\Psi (r,\theta ,\varphi ,t)=\sum\limits_{lm} {\,Y_{lm}(\theta 
,\varphi )\,}\psi _{lm}(r,t)\;.$$
For brevity, we shall omit the indices $l$,$m$, and denote $\psi 
_{lm}$
 by $\psi $. We then use the late-time expansion:
\begin{eqnarray}
\psi =\sum\limits_{k=0}^\infty  {\psi _k(r)t^{-n-k}}\;.
\label{2}
\end{eqnarray}
Here, like in Ref. \cite{1}, when studying the asymptotic behavior 
at the CH, we restrict attention to the leading-order of $\psi$
 (and $\psi _k$) in $\delta r\equiv r-r_-$.
 At this order, the functions $\psi _k$
 are polynomials in $r^*$ \cite{1}, of order $k$: \cite{6}
\begin{eqnarray}
\psi _k\cong \sum\limits_{i=0}^k {\;c^{ki}}\,{r^*}^i\;.
\label{3}
\end{eqnarray}
From the analysis in Ref. \cite{1} it is obvious that the leading-
order term in Eq. (\ref{1}), $a_0u^{-n}+b_0v^{-n}$, 
is equal to $t^{-n\kern 1pt }F^1(w)$
 (i.e. the term $j=1$ in Eq. (29) there \cite{7}), namely,
\begin{eqnarray}
a_0u^{-n}+b_0v^{-n}=t^{-n}\,[c^{0,0}+c^{1,1}(r^*/ 
t)+c^{2,2}(r^*/ t)^2+...\,\kern 1pt ]\,\;.
\label{4}
\end{eqnarray}
Therefore, the coefficients $a_0$
 and $b_0$
 depend only on the parameters $c^{k,k}$.
 We shall now show that $a_0$
 and $b_0$
 can be determined directly from $c^{0,0}$
 and $c^{1,1}$. For simplicity, let us define
$$\hat u\equiv -u\;\,,\;\,\hat a_0\equiv (-1)^na_0\;,$$
such that  $a_0u^{-n}=\hat a_0\hat u^{-n}$.
 Recalling that $\hat u=t-r^*\;\,,\;\,v=t+r^*$, we may write 
\begin{eqnarray}
a_0u^{-n}+b_0v^{-n}=t^{-n}\left[ {\hat a_0(1-r^*/ 
t)^{-n}+b_0(1+r^*/ t)^{-n}} \right] \nonumber \\
=t^{-n}\left[ (\hat a_0+b_0)+n(\hat a_0-b_0)\,(r^*/ t)
+{n(n+1)(\hat a_0+b_0) \over 2}\,(r^*/t)^2+... \right]\;.
\label{5}
\end{eqnarray}
Comparing Eqs. (\ref{4}) and (\ref{5}), we find 
\begin{eqnarray}
c^{0,0}=\hat a_0+b_0\quad ,\quad c^{1,1}=n(\hat a_0-b_0)\;.
\label{6}
\end{eqnarray}
Therefore, all we need to do is to calculate the coefficients 
$c^{0,0}$
 and $c^{1,1}$.
 This, in turn, requires the solution of the differential 
equation for $\psi _{k=0}$
 and $\psi _{k=1}$.
 Recall that for that purpose it is not sufficient to 
consider the leading order of these functions in $\delta r$:
 Since the parameters $c^{0,0}$
 and $c^{1,1}$
 will be determined by matching the functions $\psi _0$
 and $\psi _1$
 to the initial data at the EH, we shall need the exact form of 
these functions in the entire range $r_-\kern 1pt <r<r_+$.
	
The two functions $\psi _0$
 and $\psi _1$
 satisfy the same "static" (i.e. t-independent) KG equation,
$$f\psi _{k,rr}+\left( {f_{,r}+2f/ r} \right)\psi _{k,r}-
{{l(l+1)} \over {r^2}}\psi _k=\,\,0\;\;\;\;\;\;(k=0,1)\;,$$
where $f\equiv 1-2M/ r+e^2/ r^2$. 
 The general solution of this equation is \cite{8}
$$\psi _k=a\,Q_{l\,}\left( x \right)+b\,P_{l\,}\left( x 
\right)\;,$$
where $a$ and $b$ are arbitrary constants, $P_l$
 and $Q_l$
 are the Legendre functions of the first and second kinds, 
respectively, and
$$x\equiv {{2r-r_+-r_-} \over {r_+-r_-}}\;.$$
Thus, the most general expressions for $\psi _0$
 and $\psi _1$
 are
$$\psi _0=a^0\,Q_{l\,}\left( x \right)+b^0\,P_{l\,}\left( x 
\right)$$
	
and
\begin{eqnarray}
\psi _1=a^1\,Q_{l\,}\left( x \right)+b^1\,P_{l\,}\left( x 
\right)\;.
\label{7}
\end{eqnarray}
The parameters $a^0$, $b^0$, $a^1$
 and $b^1$
 are to be determined from the initial conditions at the EH. 
The event and inner horizons correspond to $x=1$ and $x=-1$, 
respectively. At both points, $P_l(x)$
 is regular and $Q_l(x)$
 diverges logarithmically. The asymptotic form of the two 
Legendre functions at the horizons is given by
\begin{eqnarray}
P_{l\,}(x\kern 1pt =\kern 1pt 1)=1\quad ,\quad Q_l(x)=-{1 \over 2}
\ln (1-x)+regular\;term\quad \,\,(x\to 1)
\label{8}
\end{eqnarray}
and
\begin{eqnarray}
P_{l\,}(x\kern 1pt =-1)=(-1)^l\quad ,\quad Q_l(x)={(-
1)^l\over 2}\ln (1+x)+\;regular\;term\quad \,\,(x\to -1)\;.
\label{9}
\end{eqnarray}
As was shown in Ref. \cite{1}, regularity at the EH implies $a^0=0$, so
\begin{eqnarray}
\psi _0=b^0\,P_{l\,}\left( x \right)\;.
\label{10}
\end{eqnarray}
Comparing Eqs. (\ref{3}), (\ref{9}) and (\ref{10}), we find 
\begin{eqnarray}
c^{0,0}=(-1)^l\,b^0\;.
\label{11}
\end{eqnarray}
Also, since both $r^*$ and $Q_l(x)$
 diverge logarithmically at the CH, $c^{1,1}$
 must be proportional to $a^1$.
 Since at the CH  $r\,-r_-\,\propto e^{-2\kappa _-r^*}$
  and  
$$1+x={2 \over {r_+-r_-}}\,(r-r_-)\,,$$
 we have
\begin{eqnarray}
\ln (1+x)=-2\kappa _-r^*+\;regular\;term\quad \quad (x\to -1)\;.
\label{12}
\end{eqnarray}
We find from Eqs. (\ref{3},\ref{7},\ref{9},\ref{12})
\begin{eqnarray}
c^{1,1}=(-1)^{l+1}\,\kappa _-\kern 1pt a^1\;.
\label{13}
\end{eqnarray}
	
We shall now calculate $b^0$
 and $a^1$
 from the initial data at the EH. Presumably, we have there 
$\psi =v^{-n}$, namely,
\begin{eqnarray}
\psi =t^{-n}\;[1+r^*/ t]^{-n}=t^{-n}\;[1-n(r^*/ 
t)+...(\,\kern 1pt r^*/ t)^2+...]\,\;.
\label{14}
\end{eqnarray}
It is obvious from Eq. (\ref{2}) that the term proportional to $(r^*/ 
t)^k$
 in the brackets at the right-hand side will come from $\psi 
_k$. Considering first the contribution from $\psi _{k=0}$,
 we find from Eqs. (\ref{8},\ref{10},\ref{14})
\begin{eqnarray}
b^0=1\;.
\label{15}
\end{eqnarray}
Consider next the contribution from $k=1$. At the EH, too, 
$Q_l(x)$
 is proportional to $r^*$ as both diverge logarithmically. In 
analogy with the above treatment of the CH, we now have 
$$\ln (1-x)=2\kappa _+r^*+\;regular\;term\quad \quad (x\to 1)\;.$$
Equations (\ref{7}) and (\ref{8}) then yield
$$\psi _1=-\kappa _+\kern 1pt a^1\,r^*+\;regular\;term\quad 
\quad (x\to 1)\;,$$
so Eq. (\ref{14}) implies 
\begin{eqnarray}
a^1=n/ \kappa _+\;.
\label{16}
\end{eqnarray}
	
Returning to the asymptotic behavior at the CH, we find 
[cf. Eqs. (\ref{11},\ref{13},\ref{15},\ref{16})]
$$c^{0,0}=(-1)^l\quad ,\quad c^{1,1}=(-1)^{l+1}\kern 1pt \kern 1pt 
n\kern 1pt \kappa _-/ \kappa _+\;,$$
and therefore
\begin{eqnarray}
\hat a_0=(-1)^l\,\,{{\kappa _+ -\kappa _-} \over {2\kappa 
_+}}\quad ,\quad b_0=(-1)^l\,\,{{\kappa _+ +\kappa _-} \over 
{2\kappa _+}}
\label{17}
\end{eqnarray}
[cf. Eq. (\ref{6})]. Substituting $\kappa _\pm ={{r_+-r_-} \over 
{2r_\pm ^2}}$, and recalling that $\hat a_0\equiv (-1)^na_0$, 
we finally obtain the desired expression for the reflection 
and transmission coefficients:
\begin{eqnarray}
a_0=(-1)^{l+n}\,\,{{r_-^2-r_+^2} \over {2r_-^2}}\quad ,\quad 
b_0=(-1)^l\,\,{{r_-^2+r_+^2} \over {2r_-^2}}\;.
\label{18}
\end{eqnarray}
This result is the same as the one obtained by Gursel et al 
[4] by a different method (Fourier integration in the complex 
plain). \cite{9}

This research was supported in part by the Fund for the 
Promotion of Research at the Technion.


\begin{thebibliography}{99}
\bibitem{6}	Equation (24) of Ref. [1] describes a polynomial of 
order $k+1$. However, as was shown in section VII there, 
regularity at the EH implies $c^{0,1}=0$. Since $c^{1,2}$
 vanishes too, it follows from the recursion relation for $\psi 
_k$
 [Eqs. (13,22) there] that $c^{k,k+1}$
 vanishes for all $k$.
\bibitem{7}	Note that regularity at the EH implies $F^0(w)=0$,
 because $c^{k,k+1}=0$
 [6], so the leading-order term in Eq. (29) of Ref. [1] is j=1.
\bibitem{8} 	This is essentially the "$k=0$ solution" in Ref. [4] 
[cf. Eq. (6) there], except that $\psi $
 there is $r$ times our $\psi $.
\bibitem{9}	See Eqs. (10,11,15) in [4], and note the following 
differences in terminology: (i) The coefficient related to 
transmission is denoted $b_0$
 here and  $A(0)$ in Ref. [4], whereas the one related to 
reflection is denoted $a_0$
 here and  $B(0)$ in Ref. [4]; (ii) The field denoted there by 
$\psi $
 is the KG field multiplied by $r$ (whereas here $\Psi $
 is the KG field itself); (iii)  In Ref. [4] it is presumed 
that $n\equiv 2l+2$
 (which is the case if the mode has a nonvanishing static 
initial multipole moment [2]), so $n$ is always even. In our 
analysis $n$ may be any integer. (Recall that for a mode with a 
vanishing static initial multipole moment, one expects $n=2l+3$ 
[2].)
\end{thebibliography}
\end{document}